\documentclass[aip,apl,preprint]{revtex4-1}

\usepackage{graphicx}%

\begin{document}


\title{Transient heat flux shielding using thermal metamaterials} 



\author{Supradeep Narayana}
\altaffiliation{Current address: HGST, a Western Digital Company, 5601 Great Oaks Parkway, San Jose, CA 95119}
\affiliation{Rowland Institute at Harvard, Harvard University, Cambridge, MA 02142}


\author{Salvatore Savo}
\affiliation{Rowland Institute at Harvard, Harvard University, Cambridge, MA 02142}

\author{Yuki Sato}
\affiliation{Rowland Institute at Harvard, Harvard University, Cambridge, MA 02142}


\begin{abstract}
We have developed a heat shield based on a metamaterial engineering approach to shield a region from transient diffusive heat flow.  The shield is designed with a multilayered structure to prescribe the appropriate spatial profile for heat capacity, density, and thermal conductivity of the effective medium. The heat shield was experimentally compared to other isotropic materials.
\end{abstract}


\maketitle 


Recent advances in the field of metamaterials have extended beyond the manipulation of electromagnetic waves. For example, building on the non-uniqueness problem studied in the context of electrical impedance tomography,\cite{Greenleaf} devices such as cloaks, concentrators, and rotators in the fields of electrostatics,\cite{Yang,Jiang} magnetostatics,\cite{NarayanaMag,Gomory} thermostatics,\cite{NarayanaHeat} and thermodynamics\cite{Schittny,Ma} have now been experimentally investigated. In this letter, we explore the use of metamaterials to enhance the transient shielding of a region from an external temperature change. Manipulating the heat current while providing an effective combination of low thermal conductivity and high heat capacity, we experimentally demonstrate that the principle of artificial material engineering and metamaterials may be successfully carried over to the field of thermal management and with further work may lead to practical applications.

\begin{figure}
\includegraphics{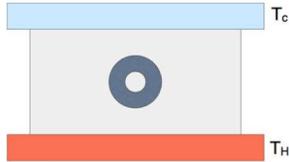}
\caption{\label{Fig1} Experimental schematic showing the shield embedded in a host background material with a heat source and a sink.}
\end{figure}

The experimental concept is depicted in Figure \ref{Fig1}. A hollow cylindrical shield is embedded in a solid host background block. Two ends of the block are placed at two different temperatures: the top side (heat sink) at $T_C=0^{\circ}C$ and the bottom (heat source) at $T_H=40^{\circ}C$. Once the temperature gradient is verified to have stabilized, the temperature at the hot end is raised to $T_H=50^{\circ}C$. We define this instant to be t=0 and monitor the temperature at the center of the shield as the thermal gradient redistributes itself. An infrared camera is used to verify the initial thermal stabilization as well as to obtain the subsequent transient temperature maps.

An ideal heat shield would have a small thermal conductivity and a large heat capacity to resist heat flow while being able to absorb heat that is injected into it without raising its temperature significantly. Given that such an ideal material is nontrivial to find, we can attempt to emulate the characteristics of such a material using the concept of metamaterial engineering.

\begin{figure}
\includegraphics{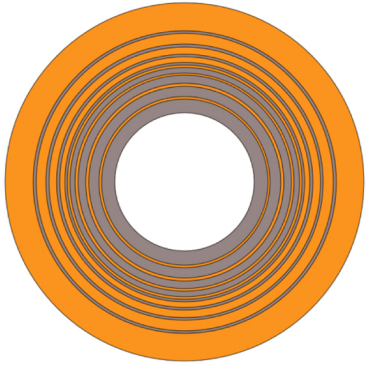}
\caption{\label{Fig2} Metamaterial thermal shield structure. The schematic shows the multilayered structures (light color: polyimide, dark color: copper) with varying thicknesses to introduce anisotropy required for heat flux guiding as well as for the desired spatial distribution of heat capacity and conductivity. The thickness ratios of the polyimide and copper layers vary from the outer solid piece to the inner piece as 10:1, 4:1, 3:1, 2:1, 1:2, 1:3, 1:4, and 1:5.}
\end{figure}

To alter and control heat flow, the effective thermal conductivity of the material has to be not only spatially varying but also direction dependent.\cite{Fan,Guenneau} Such an anisotropic material can be constructed by discretizing the overall continuous material as a network of thermal elements in series and parallel.
Viewing the overall material to be a polar grid resistor network with thermal resistances $R_r$ and $R_{\phi}$, radial and tangential components of the material conductivity $\sigma_r$ and $\sigma_{\phi}$ and the unit cell elements are related by $R_r=\Delta r/(\sigma_r r \Delta \phi h)$ and $R_\phi=r \Delta \phi/(\sigma_{\phi} \Delta r h)$, where $\Delta r$ and $\Delta \phi$ are step lengths in radial and tangential directions and $h$ represents the thickness of the material in the z direction.\cite{Yang,Jiang}
An anisotropic medium can be designed in this manner and its construction can be further simplified by the use of alternating isotropic materials as recently demonstrated both experimentally and theoretically.\cite{NarayanaHeat,Schittny,Han}

The continuity equation without a local heat source is given by $\rho C \partial T/\partial t=\bigtriangledown \cdot (\kappa \bigtriangledown T)$, where $C$ is the specific heat and $\rho$ is the material density. In the static case, the time-dependent term vanishes, and the thermal conductivities of heterogeneous composite elements become the only relevant factors. However in a dynamic scenario considered here, the spatial distribution of density and specific heat also plays a significant role.\cite{Schittny,Guenneau} A layered structure provides the flexibility to control the distribution of such parameters while achieving anisotropic conductivity required for preferential guiding of heat flux.

\begin{figure}
\includegraphics{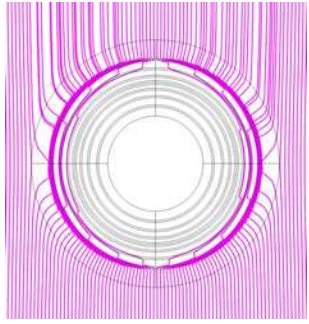}
\caption{\label{Fig3} Simulated heat flux patterns for the structure shown in Figure \ref{Fig2}.}
\end{figure}

The arrangement of the constructed metamaterial is schematically represented in Figure \ref{Fig2}. The entire composite is assembled with concentric layers of polyimide films (Fralock, Cirlex) and copper sheets of varying thicknesses. In this transient heat shield investigation, the condition that the external thermal profile exhibits negligible perturbation, often used in the cloaking experiments,\cite{NarayanaHeat,Schittny} is irrelevant. Therefore instead of enforcing strict `invisibility' with a closed cloak, the effective medium here is designed to achieve 1) a decreasing thermal conductivity towards the outside to impede heat current, 2) an increasing heat capacity towards the inner region to absorb heat that diffuses in, and 3) an anisotropic conductivity profile that reroutes heat current from radial to tangential directions. Step lengths (i.e. layer thicknesses) are chosen in order to balance these prescribed properties after parameter-sweeps with numerical simulations. We plot the simulated heat flux distribution at equilibrium in Figure \ref{Fig3}. The plot illustrates the rerouting of heat current that occurs via anisotropic conduction.

\begin{figure}
\includegraphics{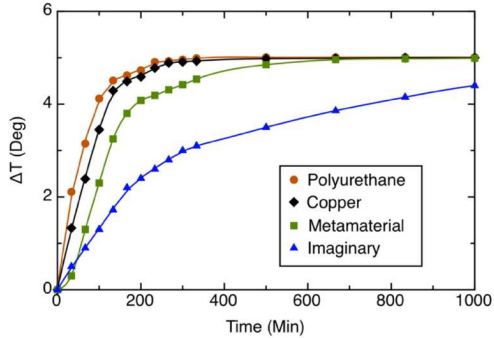}
\caption{\label{Fig4} Simulated temperature changes as a function of time at the centers of four different shields. Lines are guides to the eye.}
\end{figure}

We have first simulated the expected thermal responses for four different hollow cylindrical shields and plot the respective temperature changes at the centers of the shields as a function of time in Figure \ref{Fig4}. The four materials that we have considered are polyurethane, copper, an imaginary material with the thermal conductivity of polyurethane and specific heat and density of copper, and the thermal metamaterial shown in Figure \ref{Fig2}. Relevant material properties used for simulation are shown in Table \ref{table}. The inner and outer diameters of the shields are 2cm and 5.2cm respectively for all four cases. The parameters for the host background material are those for the silicone-based encapsulant used in the experiment described below. Simulations are performed with Comsol Multiphysics for an overall block size of 15cm(L)$\times$10cm(H)$\times$2.5cm(D). We note that this size is smaller than the one used for the experiment (20cm(L)$\times$10cm(H)$\times$5.2cm(D)) due to the restriction associated with computation time.

The polyurethane material has a low thermal conductivity and a relatively high specific heat. However due to its low density, the overall heat capacity is limited, and hence it does not constitute the most ideal shielding material in this configuration. Moreover since it is an isotropic material, there is not a way to help guide the heat flow away from the central region of interest. In the case of a copper shield, its high density makes up for the low specific heat. However its thermal conductivity is quite high. The combined effect is that, for the dimensions and the experimental configuration considered here with a well-defined heat source and a sink, the copper shield case requires more time to reach the asymptotic temperature and hence acts as a better shield than the polyurethane in this dynamic regime. Not surprisingly the imaginary material performs much better than those two shields due to its low thermal conductivity and high heat capacity.

The metamaterial shield of Figure \ref{Fig2} performs better than the two shields made of polyurethane and copper. Comparing all four materials, another interesting feature is that the initial response to the external temperature change for the metamaterial case is much slower. The prescribed thermal inertia is such that the composite even outperforms the imaginary material for short time scales. This property is a direct consequence of the nonuniform spatial profile for heat capacity and thermal conductivity in addition to the anisotropic conductivity that encourages tangential rather than radial conduction for heat current.

\begin{figure}
\includegraphics{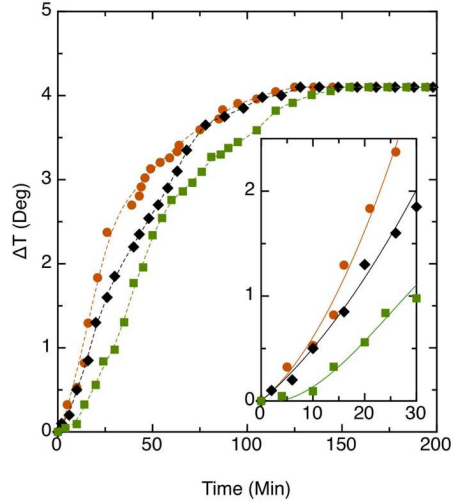}
\caption{\label{Fig5} Experimentally observed temperature changes at the shield centers. The same set of symbols are used to indicate polyurethane, copper, and metamaterial as in Fig \ref{Fig4}. The inset shows the response soon after the temperature change at the heat source.  Differences in overall characteristic relaxation times for experiment and simulation stem from the narrower dimension used for simulation, necessitated by the computation time. Dotted and solid lines are guides to the eye.}
\end{figure}

For the experiment, we have tested three hollow cylindrical materials made of polyurethane, copper, and the metamaterial (consisting of polyimide and copper). The shields are individually embedded in blocks of silicone-based thermally conductive encapsulant (Sylgard Q3-3600).  All shields have inner diameters of 2cm. The outer diameters are measured to be 5.18cm for the polyurethane, 5.2cm for copper, and 5.32cm for the metamaterial. These relative variations of $\sim3\%$ in shield diameters cause no observable differences in their thermal properties. The experimental results are plotted in Figure \ref{Fig5}. The observed temperature shifts are $\sim4.2^{\circ}C$ for all three cases, lower than the $5^{\circ}C$ expected from simulation. This is due to the fact the two end surfaces of the block in experiments are slightly higher and lower in temperatures than the actual heat sink and the source that they are respectively in contact with. As predicted in the simulation, the assembled metamaterial performs better than the other two shields made of polyurethane and copper. Although not as prominent as the simulation, the slower initial response to the external temperature change for the composite case can also be seen in the inset.

In conclusion, we have developed a heat shield based on a metamaterial engineering approach to shield a region from transient diffusive heat flow in the presence of a heat source and a sink.  The shield is designed with a multilayered structure to prescribe the desired spatial profile for heat capacity, density, and thermal conductivity of the overall medium while providing the functionality to reroute heat flux away from the region of interest. The appropriately tailored combination of materials can perform a special function better than those individual elements alone. Both the simulation and experimental results suggest that carefully tailored composite materials have the potential to act as a unique and efficient thermal shield especially if the relevant time scales of the external temperature fluctuations happen to be short. Various concepts of artificial material engineering may be successfully carried over to the field of thermal management, although the scalability in the context of effective medium approach requires further investigation for practical applications with thin films etc.

The authors thank D. Rogers for machining and S. Bevis for infrastructure support. This research was supported by the Rowland Institute at Harvard University.


%
%

%


\bibliography{Sato}

\begin{table}
\caption{\label{table}Relevant material properties.}
\begin{ruledtabular}
\begin{tabular}{llll}
Materials & Density & Specific heat & Thermal conductivity \\
 & kg/m$^3$ & J/(kg$\cdot$ K)  & W/(m $\cdot$K) \\
\hline
Polyurethane & 1050 & 1400 & 0.05 \\
Copper & 8700 & 385 & 400\\
Polyimide & 1400 & 1090 & 0.17\\
Encapsulant & 2066 & 1500 & 0.77\\
\end{tabular}
\end{ruledtabular}
\end{table}

\end{document}